\documentclass[screen, manuscript]{acmart}
\AtBeginDocument{%
  \providecommand\BibTeX{{%
    \normalfont B\kern-0.5em{\scshape i\kern-0.25em b}\kern-0.8em\TeX}}}

\copyrightyear{2022}
\acmYear{2022}
\setcopyright{rightsretained}
\acmConference[UbiComp/ISWC '22 Adjunct]{Proceedings of the 2022 ACM International Joint Conference on Pervasive and Ubiquitous Computing}{September 11--15, 2022}{Cambridge, United Kingdom}
\acmBooktitle{Proceedings of the 2022 ACM International Joint Conference on Pervasive and Ubiquitous Computing (UbiComp/ISWC '22 Adjunct), September 11--15, 2022, Cambridge, United Kingdom}\acmDOI{10.1145/3544793.3560326}
\acmISBN{978-1-4503-9423-9/22/09}

\usepackage{multirow}
\usepackage[shortlabels]{enumitem}

\usepackage[hypcap=false]{caption}
\usepackage{subcaption}
\usepackage[font=small,skip=0pt]{caption}

\begin{document}

\title{Smart Cup: An impedance sensing based fluid intake monitoring system for beverages classification and freshness detection}

\author{Mengxi Liu}
\orcid{0000-0003-0527-1208}
\author{Sizhen Bian}
\orcid{0000-0003-3723-1980}
\author{Bo Zhou}
\orcid{0000-0003-3723-1980}
\author{Agnes Grünerbl}
\orcid{0000-0002-4156-7121}
\author{Paul Lukowicz}
\orcid{0000-0003-0320-6656}
\email{FirstName.LastName@dfki.de}
\affiliation{%
  \institution{German Research Center for Artificial Intelligence (DFKI)}
  \streetaddress{Trippstadter Str. 122}
  \city{Kaiserslautern}
  \country{Germany}
  \postcode{67663}}

\renewcommand{\shortauthors}{Mengxi and Sizhen, et al.}

\begin{abstract}

This paper presents a novel beverage intake monitoring system that can accurately recognize beverage kinds and freshness. By mounting carbon electrodes on the commercial cup, the system measures the electrochemical impedance spectrum of the fluid in the cup.
We studied the frequency sensitivity of the electrochemical impedance spectrum regarding distinct beverages and the importance of features like amplitude, phase, and real and imaginary components for beverage classification. The results show that features from a low-frequency domain (100 Hz to 1000 Hz) provide more meaningful information for beverage classification than the higher frequency domain. Twenty beverages, including carbonated drinks and juices, were classified with nearly perfect accuracy using a supervised machine learning approach. The same performance was also observed in the freshness recognition, where four different kinds of milk and fruit juice were studied. 
\end{abstract}



\maketitle

\section{Introduction}

Inadequate beverages intake breaks the nutrition balance of the body and can lead to, for example, dehydration, which will cause confusion, seizures, or even death \cite{picetti2017hydration} in extreme situations. This is an issue in particular for older people. An easy-to-deploy and unobtrusive fluid intake monitoring system in daily life is a promising approach to provide individuals with qualitative and quantitative feedback and encourage them to hydrate regularly throughout the day. Researcher have explored many beverage recognition solutions \cite{lester2010automatic,haddi2014nose,rodriguez2020wpt} based on distinct sensing modalities. However, privacy issues and environmental conditions(video-based solutions), high cost(light-spectrum solutions), and measurement conditions(gas sensor solutions) are still challenges that limit the widespread use of existing solutions in everyday life.

In this paper, we designed an unobtrusive, easy-to-deploy, and low-cost  hardware system  that can be mounted in a commercial cup to automatically analyse various beverage properties. In this demo we focus on distinguishing different beverage types and freshness (where relevant) but the system can in principle also be used for other purposes such as analysing sugar concentration or coffee strength (which is currently work in progress). 

Since beverages have different electrical characteristics, beverages along with electrodes can be modeled as discrete electrical elements, like capacitors, resistors, and constant phase elements, which result in different impedance responses when supplying different frequencies of impulses to the beverages. Therefore, we analyzed the impedance spectrum of the beverages to find a frequency range where the impedance data contain more differentiable features of the beverages. Different features from the impedance spectrum were extracted as input information to a classifier. 

The proposed solution enjoys three advantages over the existing approaches: firstly, the carbon electrode implemented in our system for impedance measurement is more health-friendly, stable, and cheaper than the metal electrodes used in other solutions; Secondly, the proposed solution can recognize both the beverage kinds and freshness by a simple sensing system with unobtrusive appearance; Third, a compact machine learning model achieves nearly perfect recognition accuracy, which enables the local inference on an embedded device with constrained resources in the future.

\section{Hardware Implementation}

Figure \ref{fig:hwarchitecture} shows the hardware architecture and Figure \ref{fig:prototype} shows the prototype of the beverage recognition system. Four carbon electrodes with 2mm diameter and 10 mm length each were customized as the sensing units and mounted at the bottom of the cup. The analog front-end chip AD5940 for impedance measurement was connected to the carbon electrodes. Data was collected by the microcontroller ADUCM3029.

\begin{minipage}{\textwidth}
    \begin{minipage}[b]{0.55\textwidth}
         \begin{minipage}[b]{0.68\textwidth}
             \centering
             \includegraphics[width=\textwidth]{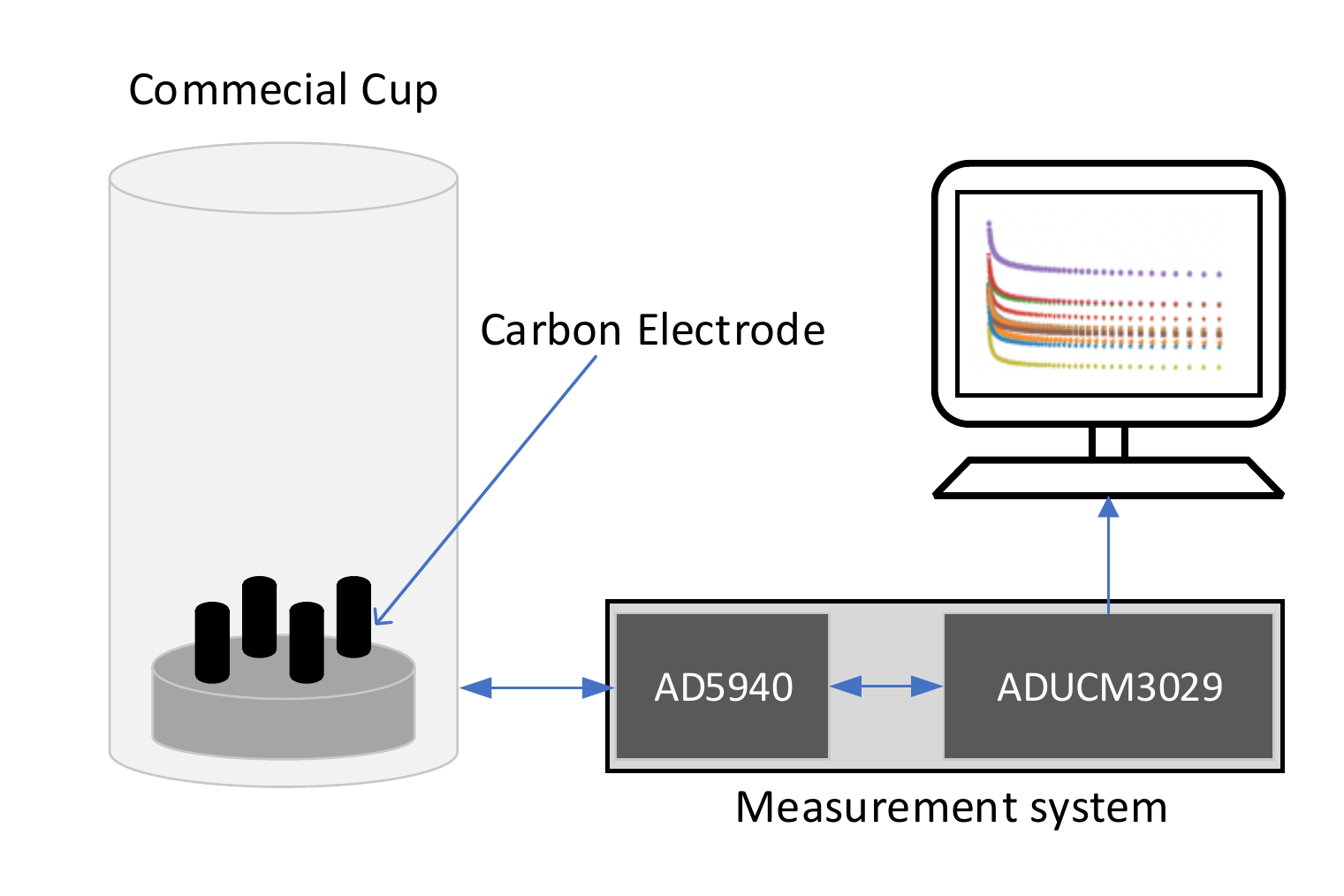}
             \captionof{figure}{Hardware architecture}
             \label{fig:hwarchitecture}
             \vspace{-8mm}
         \end{minipage}
         \hfill
         \begin{minipage}[b]{0.3\textwidth}
             \centering
             \includegraphics[width=\textwidth]{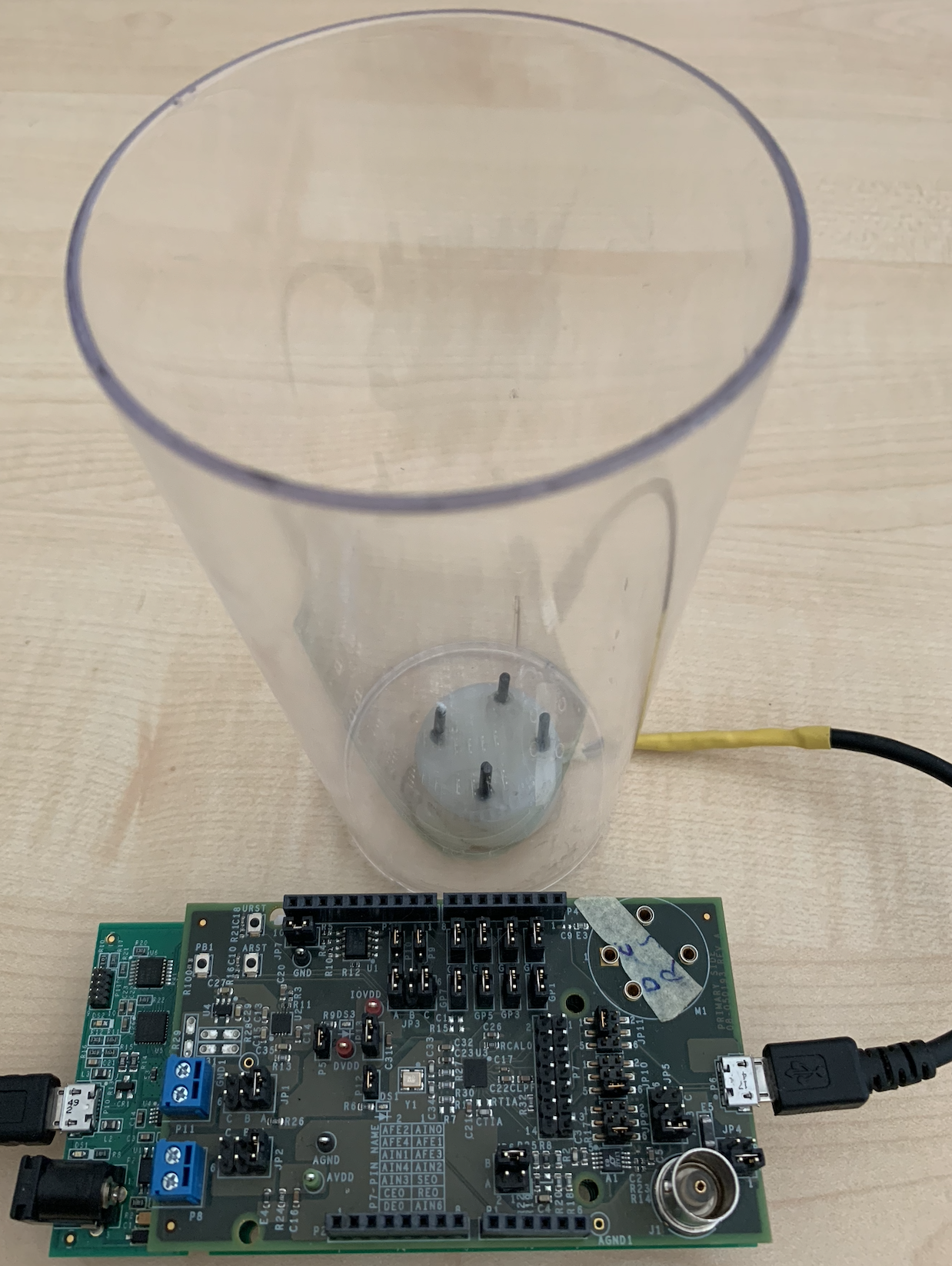}
             \captionof{figure}{Prototype}
             \label{fig:prototype}
             \vspace{-8mm}
         \end{minipage}
    \end{minipage}
    \hfill
    \begin{minipage}[b]{0.43\textwidth}
        \renewcommand{\arraystretch}{0.7}
        \resizebox{0.93\textwidth}{!}{
        \captionsetup{type=table}
        \begin{tabular}{|c|c|c|c|}
            \hline
            No.&Beverage&No.&Beverage\\
            \hline
            0&Mineral water&10& Colar classic 2\\
            \hline
            1&Cola Zero 1&11&Cola Zero 3\\
            \hline
            2&Orange Zero&12&Eistee Pfirsch\\
            \hline
            3&Cola Light&13&Apfel Schorle\\
            \hline
            4&Cola Mix&14&Banana juice\\
            \hline
            5&Cola Classic 1&15&Pineapple juice\\
            \hline
            6&Cola Zero 2&16&Currants juice\\
            \hline
            7&Sprite&17&Orange juice\\
            \hline
            8&7 UP&18&Carrots juice\\
            \hline
            9&Fanta&19&Mixed vegetable juice\\
            \hline
        \end{tabular}}
        \captionof{table}{Beverage information}
        \label{tab:beverage}
        \vspace{-8mm}
    \end{minipage}
\end{minipage}

\section{Evaluation}
\subsection{Beverage Kind Classification}
\begin{figure}
     \centering
     \begin{subfigure}[b]{0.45\textwidth}
         \centering
         \includegraphics[width=\textwidth]{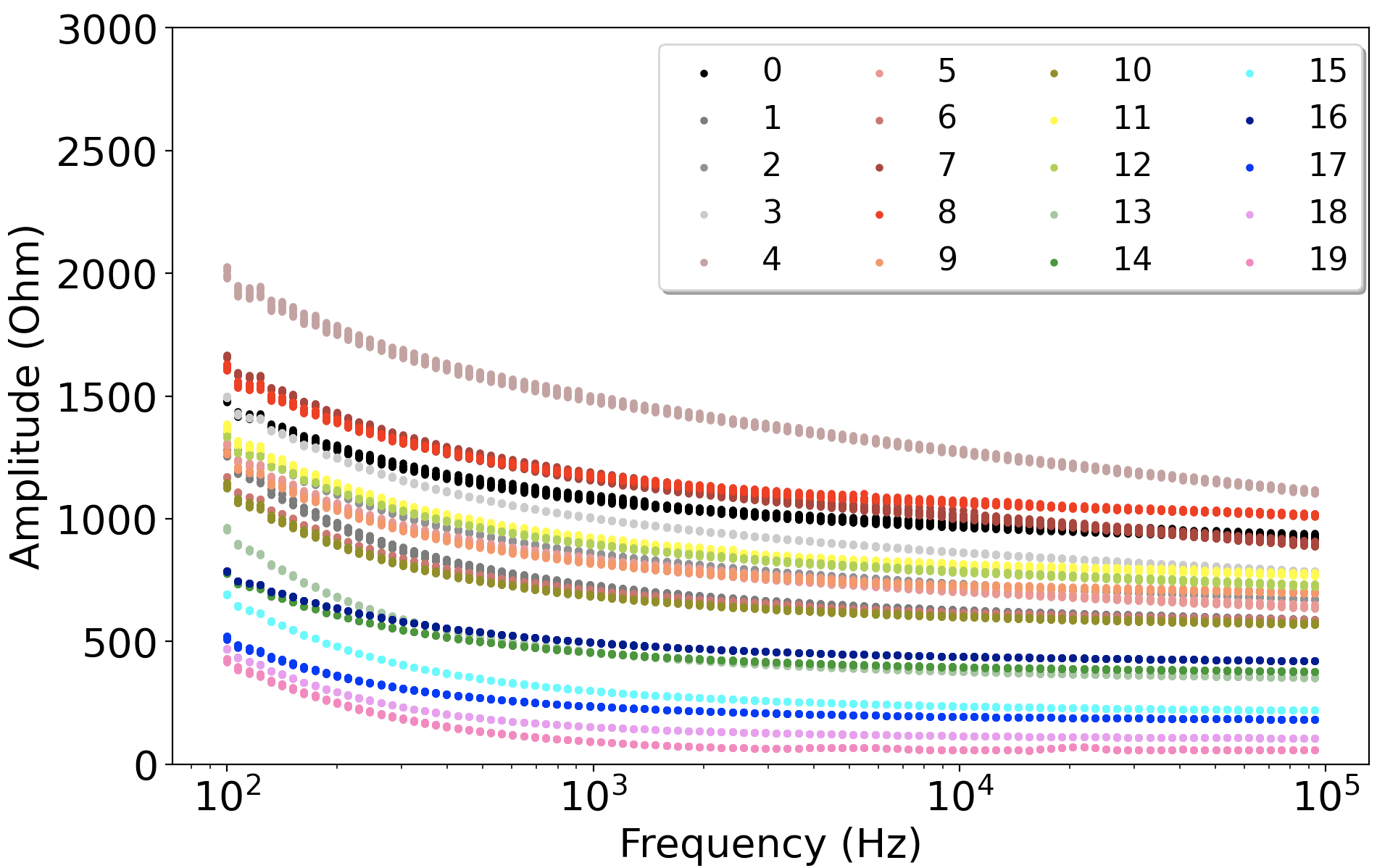}
         \caption{}
         \label{rawdata_af}
     \end{subfigure}
     \hfill
     \begin{subfigure}[b]{0.45\textwidth}
         \centering
         \includegraphics[width=\textwidth]{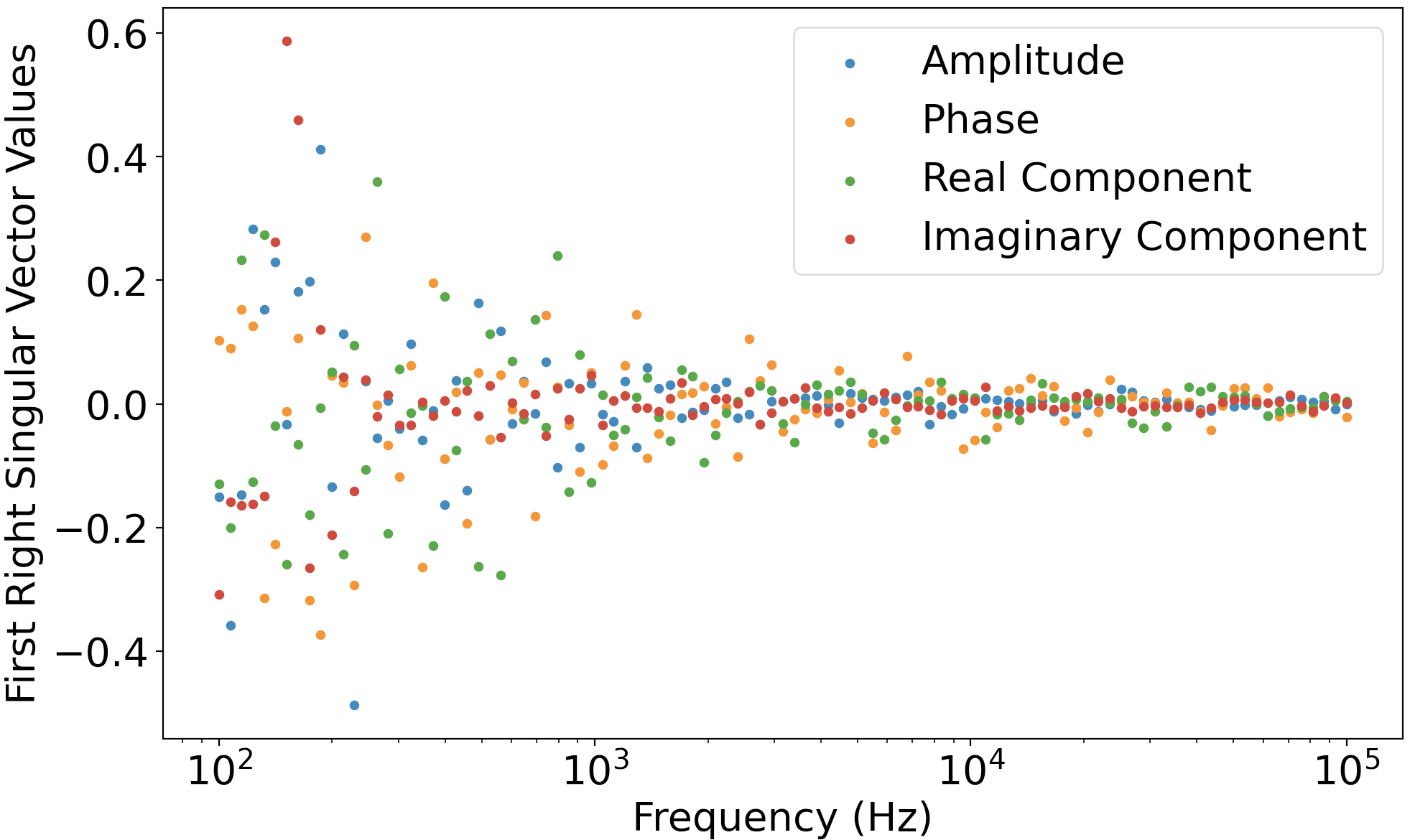}
         \caption{}
         \label{svd_ap}
     \end{subfigure}
        \caption{a). Amplitude and frequency diagram of 20 kinds of beverages. b). First right singular vector values from amplitude, phase, imaginary and real component. 1. Most meaningful amplitude : 245.47 Hz. 2. Most meaningful phase : 162.18 Hz. 3. Most meaningful real component : 263.03 Hz. 4. Most meaningful imaginary component : 151.36 Hz.}
        \label{combined_plot}
        \vspace{-5mm}
\end{figure}

In the first set of experiments, twenty beverages as shown in Table \ref{tab:beverage}, including carbonated drinks and juice, were tested. Each kind of beverage was sampled ten times. The stimuli signal is a sinusoidal signal with an amplitude of 50 mV, the response signals of 101 frequency points from 100 Hz to 100000 Hz were measured. The dataset consisted of 200 observations, including amplitude and phase information of beverages. Figure \ref{rawdata_af} plots the amplitudes feature of the dataset. All the experiment is conducted in a daily living environment under room temperature. The measured beverage temperature is within 20 to 23 \textdegree C.

At first, we utilized Singular Value Decomposition (SVD) to find a smaller
frequency domain where the features contain the most meaningful information of each beverage to reduce feature dimensions. Figure \ref{svd_ap} shows the first right singular vector values from these features, the most influential features are located in the low frequency region under 1000 Hz. Therefore, we extracted a reduced feature set that only includes 20 frequency points from 100 Hz to 1000 Hz.

We then used Random Forest (RF) and Dense Neural Net (DNN) to classify the beverages. The summary of the beverage kinds classification result is listed in Table \ref{tab:classification}, where the recognition accuracy of beverage kinds is nearly perfect.

\begin{minipage}{\textwidth}
    \begin{minipage}{0.55\textwidth}
        \renewcommand{\arraystretch}{0.8}
        \resizebox{1\textwidth}{!}{
        \captionsetup{type=table}
        \begin{tabular}{|c|c|c|c|c|c|c|}
        \hline
        Classifier Model& RF & RF & RF & RF & DNN & DNN\\
        \hline
        Dataset&  A &  B &  C &D &C&  D\\
        \hline
        Full features & 1.0  & 1.0 & 1.0 & 1.0 & 1.0& 0.98\\
        \hline
        Reduced features & 0.98 &0.99 & 0.95 &0.98 &0.93& 0.98\\
        \hline
        \end{tabular}}
        \captionof{table}{Summary of Beverage Kinds Classification Results (Dataset A includes real and imaginary component. Dataset B includes amplitude and phase. Dataset C only includes amplitude feature. Dataset D only includes phase feature)}
        \label{tab:classification}
        \vspace{-3mm}
    \end{minipage}
    \hfill
    \begin{minipage}{0.42\textwidth}
    \renewcommand{\arraystretch}{1}
        \resizebox{\textwidth}{!}{
        \captionsetup{type=table}
        \begin{tabular}{|c|c|c|c|c|}
        \hline
        Beverage & Milk 1 & Milk 2 & Juice 1& Juice 2\\
        \hline
        Amplitude&  1.0 &  1.0 &  1.0 & 1.0 \\
        \hline
        Phase & 1.0 & 1.0 & 1.0 & 1.0\\
        \hline
        \end{tabular}}
        \captionof{table}{Summary of Beverage Freshness Classification Results by Random Forest Classifier}
        \label{tab:freshnessClassification}
    \end{minipage}
\end{minipage}

\subsection{Beverage Freshness Detection}
We tested the freshness of two kinds of milk and two kinds of juice. The data collection procedure was the same as in the previous classification experiments. Each dataset consisted of 30 observations, three freshness classes (0 hours, 24 hours, and 48 hours since unpacked), and 202 features (101 amplitudes, 101 phases). Table \ref{tab:freshnessClassification} shows the freshness classification results of the four kinds of beverages. The accuracy of freshness classification of each dataset is up to 100 \%, with features from both amplitude and phase.

\section{Conclusion and future work}
This work demonstrated a novel automatic beverage intake monitoring system with the carbon electrodes-based impedance sensing unit. Twenty beverages, including carbonated drinks and juice in daily life, were included for beverage classification, and four beverages, including milk and juice, were tested for freshness detection. The results of both experiments with nearly perfect accuracy showed that our proposed system with unobtrusive appearance and low cost has immense potential as an automatic beverage monitoring system in daily life. In the future, we will investigate monitoring more general beverage information by this proposed system, like the chemical composition analysis. 
\bibliographystyle{ACM-Reference-Format}
\bibliography{sample-base}

\end{document}